\documentclass[onecolumn,showpacs,preprintnumbers,amsmath,amssymb,aps,pra,groupedaddress]{revtex4-2}
\usepackage{multirow,diagbox}  
\usepackage{siunitx}
\usepackage[colorlinks,citecolor=blue,linkcolor=red,hyperindex,CJKbookmarks]{hyperref}
\usepackage{amsmath}
\usepackage{graphicx}
\usepackage{dcolumn}
\usepackage{subfigure}
\usepackage{mathrsfs}
\usepackage{amssymb}
\usepackage{amsfonts}
\usepackage{makecell}
\usepackage{color}
\begin{document}
\draft
\title{Comment on "Quantum coherence between mass eigenstates of a neutrino
cannot be destroyed by its mass-momentum entanglement"}
\author{Shi-Biao Zheng}\thanks{%
E-mail: t96034@fzu.edu.cn}
\address{College of Physics and Information Engineering, Fuzhou University,\\
Fuzhou 350108, China}

\begin{abstract}
In arXiv:2410.21850, I proved that the quantum coherence between the mass
eigenstates of a neutrino will be destroyed if they are correlated with
different momenta. In arXiv:2411.01190, James M. Cline claimed that I had
made the unrealistic assumption that the neutrino is always in a nearly
exact energy eigenstate, and ignored the spatial dependence of the
wavefunction in my paper. However, I did not assume that the neutrino is in
a nearly exact eigenstate of energy anywhere in my paper, and the
wavefunction I wrote in the position representation has a spatial
dependence. The argumentation of arXiv:2411.01190 is based on
misinterpreting my claim, and on ignoring the critical fact that the
neutrino's wavepacket has a finite size and the detector has a large volume.
\end{abstract}

\vskip0.5cm

\narrowtext
\maketitle
\bigskip In Ref. 1, I proved that the quantum coherence between the mass
eigenstates of a neutrino will be destroyed if they are correlated with
different momenta. This point can be well understood in the momentum
representation, where the state of the neutrino can be written as 
\begin{equation}
\left\vert \nu \right\rangle =%
\mathop{\displaystyle\sum}%
\limits_{j}\int_{{\bf \sigma }_{j}}d^{3}{\bf p}_{j}f({\bf p}_{j})\left\vert 
{\bf p}_{j}\right\rangle \left\vert \nu _{j}\right\rangle .
\end{equation}%
Here $\left\vert \nu _{j}\right\rangle $ denotes the $j$th mass eigemstate,
and ${\bf \sigma }_{j}$ represents the distribution region of the momentum
associated with the mass eigenstate $\left\vert \nu _{j}\right\rangle $,
with the probability amplitude distribution function $f({\bf p}_{j})$. As
the flavor oscillations are assumed to arise from the quantum coherence
among the mass eigenstates, the momentum degree of freedom needs to be
traced out for the discussion of such oscillations. This leaves the mass
degree of freedom in a classical mixture, given by the density operator 
\begin{equation}
\rho =%
\mathop{\displaystyle\sum}%
\limits_{j,k}D_{j,k}\left\vert \nu _{j}\right\rangle \left\langle \nu
_{k}\right\vert ,
\end{equation}%
where 
\begin{eqnarray}
D_{j,k} &=&\int d^{3}{\bf p}\int_{{\bf \sigma }_{j}}\int_{{\bf \sigma }%
_{k}}f({\bf p}_{j})f^{\ast }({\bf p}_{k})\left\langle {\bf p}\right\vert
\left. {\bf p}_{j}\right\rangle \left\langle {\bf p}_{k}\right\vert \left. 
{\bf p}\right\rangle   \nonumber \\
&=&\int_{{\bf \sigma }_{j}}\int_{{\bf \sigma }_{k}}f({\bf p}_{j})f^{\ast }(%
{\bf p}_{k})\left\langle {\bf p}_{k}\right\vert \left. {\bf p}%
_{j}\right\rangle .
\end{eqnarray}%
When there is no overlapping between ${\bf \sigma }_{j}$ and ${\bf \sigma }%
_{k}$, we have ${\bf p}_{j}\neq {\bf p}_{k}$, which leads to $D_{j,k}=0$ for 
$j\neq k$. This implies that the coherence among the mass eigenstates is
destroyed when they are entangled with different momenta, so that the flavor
oscillations cannot occur.

The entanglement-induced decoherence can also be understood in terms of
quantum-mechanical complementarity [3,4]. When the mass eigenstates are
correlated with different momenta, one can know in which mass eigenstate the
neutrino is by measuring its momentum in principle. This is sufficient to
destroy the coherence among the mass eigenstates. It does not matter that
the momentum is not measured actually [5]. Such a complementary principle,
which has passed a number of experimental tests [5-11], cannot be violated.

In the above derivation, there is no restriction on the energy and momentum
for each mass eigenstate. In Ref. 2, James M. Cline seriously misinterpreted
my starting point by saying "{\sl The essence of the claim is that a flavor
state emitted from a weak interaction will be in a nearly exact eigenstate
of energy}." As a matter of fact, I have assumed that the momentum
associated with each mass eigenstate has a spread. When the spread is
sufficiently large, the neutrino cannot be in a nearly exact eigenstate of
energy. The statement that the spatial dependence of the neutrino
wavefunction was ignored in my derivation is also incorrect. 

For simplicity, we here suppose that the neutrino travels along the z
direction. Then, in the position representation, the evolution of the wave
function is given by%
\begin{equation}
\left\vert \varphi _{\nu }(z,t)\right\rangle =%
\mathop{\displaystyle\sum}%
\limits_{j}g_{j}(z,t)\left\vert \nu _{j}\right\rangle ,
\end{equation}%
where 
\begin{equation}
g_{j}(z,t)=(2\pi )^{-1/2}\int_{{\bf \sigma }_{j}}dp_{j}f(p_{j})e^{i(p_{j}z%
{\bf -}E_{j}t)},
\end{equation}%
and $E_{j}=\sqrt{p_{j}^{2}+m_{j}^{2}}$ with $m_{j}$ being the mass of the $j$%
th mass eigenstate. Contrary to what was said in Ref. 2, the function $%
g_{j}(z,t)$ has a spatial dependence. For a definite position $z$, the
coherence between $\left\vert \nu _{j}\right\rangle $ and $\left\vert \nu
_{k}\right\rangle $ is manifested by the relative phase factor $%
e^{i(p_{j}-p_{k})z}$. However, the neutrino detector has a large volume, and
it was not recorded at which point of the detector the detected neutrino
reacts with the medium in neutrino experiments. The probability for
detecting the electron flavor on the detector is given by 
\begin{equation}
P_{e}=\int_{D}dz\left\vert \left\langle \nu _{e}\right\vert \left. \varphi
_{\nu }(z,t)\right\rangle \right\vert ^{2}
\end{equation}%
where 
\begin{equation}
\left\vert \nu _{e}\right\rangle =%
\mathop{\displaystyle\sum}%
\limits_{j}U_{ej}\left\vert \nu _{j}\right\rangle 
\end{equation}%
is the electron flavor eigenstate, and $D$ is the detection region.
Substituting Eqs. (4) and (7) into (6), we obtain%
\begin{equation}
P_{e}=(2\pi )^{-1}%
\mathop{\displaystyle\sum}%
\limits_{j,k}U_{ej}^{\ast }U_{ek}\int_{D}dz\int_{{\bf \sigma }%
_{j}}dp_{j}\int_{{\bf \sigma }_{k}}dp_{k}f(p_{j})f^{\ast
}(p_{k})e^{i(p_{j}-p_{k})z}e^{i(E_{k}{\bf -}E_{j})t}.
\end{equation}%
When the neutrino detector has a size much larger than the wavepacket size,
it is reasonable to replace $\int_{D}dze^{i(p_{j}-p_{k})z}$ with $%
\int_{-\infty }^{\infty }dze^{i(p_{j}-p_{k})z}$. For $p_{j}\neq p_{k}$, this
integral vanishes, and $P_{e}$ is approximated by%
\begin{equation}
P_{e}\simeq 
\mathop{\displaystyle\sum}%
\limits_{j}\left\vert U_{ej}\right\vert ^{2},
\end{equation}%
which does not show interference effects. This is due to the fact that the
position-dependent phase factor $e^{i(p_{j}-p_{k})z}$, which is responsible
for the spatial interference, is averaged out when the size of the detection
region is larger than that of the neutrino wavepacket. The argumentation of
Ref. 2 is valid only when the wavepacket size of the neutrino is much larger
than the detector size, which does not coincide with neutrino experiments,
where the detector has a large volume.

In Ref. [12], I further pointed out that the 2/3 deficit of solar $^{8}$B
electron neutrinos cannot be reasonably interpreted in terms of the
inconsistency between the flavor and mass eigenstates even if the
corresponding superposition of mass eigenstates can be produced. This is due
to the fact that the states of a neutrino and an electron are entangled
after their charged-current interaction, which has been ignored in previous
investigations of the matter effect [13-15]. Due to this nonseparability,
the effects of the electrons cannot be modeled as a static potential for
solar $^{8}$B neutrinos. Consequently, during the propagation $^{8}$B
neutrinos cannot adiabatically evolve to a pure mass eigenstate, which was
supposed to have an 1/3 overlapping with the electron flavor. In Ref. 12, I
proposed an alternative mechanism, where neutrino oscillations are caused by
virtual excitation of the Z bosonic field, which can connect different
neutrino flavors. Under the competition between the coherent coupling
induced by the Z bosonic field and the decoherence effect caused by
charged-current interactions, solar $^{8}$B neutrinos would finally evolve
to a steady state, where the electron flavor has a 1/3 population.

Finally, it should be pointed out that it is unreasonable to use the
Standard Model of particle physics to criticize the mechanism proposed in
Ref. 12. It is a well known fact that neutrino oscillations themselves are a
phenomenon that is beyond the Standard Model, and certainly cannot be
interpreted in the framework of the Standard Model. In other words, if one
insists to judge the correctness of an interpretation in terms of the
Standard Model, then there does not exist any correct interpretation.

\end{document}